\documentclass[aps,prb,reprint,superscriptaddress,longbibliography]{revtex4-2}

\usepackage[utf8]{inputenc}
\usepackage{amsmath}
\usepackage{amssymb}
\usepackage{braket}
\usepackage{graphicx}
\usepackage{mathtools}
\usepackage[nice]{nicefrac}
\usepackage{dsfont}
\usepackage[dvipsnames]{xcolor}
\DeclareMathOperator{\sgn}{sgn}

\begin{document}

\title{Quantum dot coupled to topological insulators: The role of edge states}

\author{M.~T.~Maurer}
\affiliation{Institut f{\"u}r Theorie der Statistischen Physik, RWTH Aachen University and
  JARA---Fundamentals of Future Information Technology, 52056 Aachen, Germany}
\author{Y.-T.~Lin}
\affiliation{Institut f{\"u}r Theorie der Statistischen Physik, RWTH Aachen University and
  JARA---Fundamentals of Future Information Technology, 52056 Aachen, Germany}
\author{D.~M.~Kennes}
\affiliation{Institut f{\"u}r Theorie der Statistischen Physik, RWTH Aachen University and
  JARA---Fundamentals of Future Information Technology, 52056 Aachen, Germany}
 \affiliation{Max Planck Institute for the Structure and Dynamics of Matter, Center for Free Electron Laser Science, 22761 Hamburg, Germany}
\author{M.~Pletyukhov}
\affiliation{Institut f{\"u}r Theorie der Statistischen Physik, RWTH Aachen University and
  JARA---Fundamentals of Future Information Technology, 52056 Aachen, Germany}
\author{H.~Schoeller}
\affiliation{Institut f{\"u}r Theorie der Statistischen Physik, RWTH Aachen University and
  JARA---Fundamentals of Future Information Technology, 52056 Aachen, Germany}
\author{V.~Meden}
\affiliation{Institut f{\"u}r Theorie der Statistischen Physik, RWTH Aachen University and
  JARA---Fundamentals of Future Information Technology, 52056 Aachen, Germany}

\begin{abstract}
We investigate a system consisting of one or two topological-insulator leads which are tunnel coupled to a single dot level. The leads are described by the one-dimensional Su-Schrieffer-Heeger model. We show that (topological) edge states cause characteristic features in the dot spectral function, the dot occupation, and the finite-bias current across the dot.  
As the kinetic energy is quenched in the dot region, local two-particle interactions are of particular relevance there. This motivates us to test whether the aforementioned edge-state features are robust against such interactions; we report here that they are either robust or even enhanced.
We conclude that the characteristic features can be used to determine if the leads are in their topologically non-trivial or trivial phase.
\end{abstract}

\date{\today}

\maketitle

\section{Introduction}
\label{sec:intro}

Today, the topic of  topological phases of matter is omnipresent in condensed matter physics research, see, e.g., Refs.~\onlinecite{Hasan10,Qi11,Asboth16}.
A topological insulator is gapped, similar to a normal band insulator, but is further characterized by topological invariants related to bulk properties. Via the bulk-boundary correspondence \cite{Fidkowski11,Mong11,Gurarie11,Essin11,Fukui12,Yu17,Rhim18,Silveirinha19}, these invariants are linked to topological edge states localized at boundaries or interfaces. A major focus of research is the identification of the signatures of such states in theoretical studies and their experimental detection. A crucial question is how topological edge states can be distinguished from other in-gap states localized at boundaries or interfaces. Most prominently, perhaps, this has been pursued in the case of Majorana bound states in topological superconductors \cite{Prada20}, valued due to their promise in topological quantum computing applications \cite{Nayak08,Aasen16}. 

A prototypical realization of a topological insulator is the chiral, one-dimensional Su-Schrieffer-Heeger (SSH) model, originally conceived to study the conductivity of doped polyacetylene \cite{Chiang77, Shirakawa77, Su79, Heeger88, Heeger01}. It is a tight-binding chain with staggered hopping parameters $t_1$ and $t_2$ that, depending on the ratio of the two hopping matrix elements, may host a localized edge state at the interface to a topologically trivial material, e.g., the vacuum. Apart from the original polymer system, the SSH model was realized experimentally and the topological edge states were detected in more controllable settings such as cold atoms \cite{Leder16,Meier16,Giergiel19,Leseleuc19}, electronic states in artificial atomic lattices \cite{Drost17} or superlattices \cite{Belopolski17}, mechanical chains \cite{Chaunsali17}, and photonic systems \cite{Kitagawa12,Zeuner15,Bleckmann17}.
Among theoretically predicted signatures of the topological edge states are effects in the entanglement entropy \cite{Micallo20}, in the decoherence of a coupled qubit \cite{Zaimi21}, and in transport and noise characteristics in non-equilibrium settings where a current is driven through an SSH chain \cite{Niklas16,Bello16,Benito16,Bohling18,Balabanov20}. However, to the best of our knowledge, the archetypal mesoscopic setup of a quantum dot coupled to leads has not yet been studied if the latter are modeled as SSH chains.

The aim of the present paper is to study such a setup. In particular, we investigate the effects of edge states on the occupation as well as on the spectral and on the transport properties of a single-level quantum dot that is weakly tunnel coupled to the boundary of one or two semi-infinite fermionic SSH chains. In our study of the single-lead (equilibrium) setup, we complement the results of Ref.\ \onlinecite{Zaimi21} regarding edge-state detection via a coupled dot.

This article is divided into two parts. Initially, we consider the fermions to be non-interacting. This allows us to employ the Green's function formalism in order to obtain exact results for the equilibrium dot spectral function and the dot occupation as well as the steady-state current flowing through the dot in a two-terminal system with a bias voltage applied to the SSH leads.
We identify signatures of the topological edge states in the dot spectral function and interpret them in terms of an effective two-state model that takes into account the hybridization and repulsion of the dot and the edge state. These signatures translate to effects in the dot occupation and the current. 

In equilibrium, the occupation of the dot as a function of its energy is strongly broadened by the presence of the edge state. While the curve shows a sharp step when the lead is in its normal phase at zero temperature, the step acquires a width of $\sqrt{\Gamma \Delta}$ when the lead is topological. Here, $\Gamma$ characterizes the tunneling between the dot and the SSH lead, and $\Delta$ is the energy gap in the lead. In a non-equilibrium setup with two SSH leads and a bias voltage applied across the dot,
the presence of edge states leads to a shift of the current-voltage curve, the sign of which (retardation vs.\ early current onset) depends on the phases of both leads and further system parameters such as the dot energy.

In the second part of the paper, we consider a local two-particle interaction $U$ in the dot region. In this part of our setup, the two-particle interaction is particularly relevant because of the quenched kinetic energy (tunnel coupling). Employing first-order perturbation theory in $U$, we show that the aforementioned signatures of edge states are robust against this perturbation. For the setup and parameters of interest to us (see below), we can ignore correlation effects typical for quantum dots, such as the Kondo effect \cite{Hewsonbook} or the physics of the interacting resonant level model, see, e.g., Ref.~\onlinecite{Karrasch10}. Therefore, we do not have to use more elaborate quantum many-body methods.

We conclude that the measurement of the dot occupation constitutes a simple method to detect an edge state and thus to determine if the SSH chain is in its topologically non-trivial phase. A single measurement suffices to this end, as the presence or absence of any broadening (at zero temperature) directly reveals the lead's phase.
The shift of the current-voltage curve, on the other hand, can only serve to detect a phase change of a lead, as the form of the curve itself is unaltered. Nonetheless, we report it here as a straightforward signature of edge states in transport experiments, which can be explained in a simple two-state picture of the dot and the edge state.

\section{The model}
\label{sec:model}

\begin{figure}[t]
\includegraphics[width=0.475\textwidth]{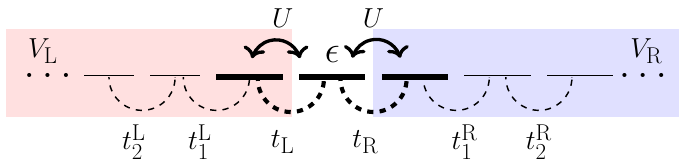}
\caption{Sketch of the single quantum-dot level coupled to two semi-infinite Su-Schrieffer-Heeger chains.}
\label{Fig:sketch}
\end{figure}

We consider spinless fermions in a system consisting of a single quantum-dot level with onsite energy $\epsilon$, described by $H_{\rm dot} = \epsilon \, d^{\dagger} d$ (standard second-quantized notation), that is weakly tunnel coupled to SSH leads. In solid-state setups, the fermions would be electrons whose spin degree is polarized by the application of a Zeeman field large enough to cause an energy splitting larger than the band width of the leads.
The total system is given by the Hamiltonian $H = H_{\rm dot} + H_{\rm leads} + H_{\rm t} + H_{\rm int}$; see also Fig.\ \ref{Fig:sketch}.

We model the leads as SSH chains of non-interacting fermions, $H_{\rm leads} = \sum_{l}H_{l}$, with $H_{l} = \sum_{j=1}^{L_l} \big[-t^{l}_1\, a_{l, 2j}^{\dagger}\,a_{l,2j-1}\, - \,t^{l}_2\, a_{l, 2j + 1}^{\dagger}\, a_{l,2j} \,+\, {\rm H.c.}\big] + V_{l} N_{l}$, where $t_1^l, t_2^l>0$. Here $l$ labels the leads, $j$ the unit cells, $L_l$ is the number of unit cells, $N_l$ denotes the particle number operator in lead $l$, and $a_{l,n}$ destroys a fermion on the $n$-th site of lead $l$. 
Furthermore, we have set (here and in the following) the charge $e=1$, similarly $\hbar = 1$. 
At the end of any calculation, we take the thermodynamic limit $L_l \to \infty$ (semi-infinite chain) to ensure that the leads act as infinite reservoirs. We exclusively focus on the zero temperature limit. For the case of two leads we use $l = \mbox{L/R}$ for the left and right one, respectively. By choosing different bias-voltage parameters $V_{\rm L}$ and $V_{\rm R}$, we can tune the system to non-equilibrium and drive a current through the dot. The energy spectrum of SSH chain $l$ is symmetric around $V_{l}$ with an energy gap of $2\Delta_{l} = 2|t^{l}_1 - t^{l}_2|$ and bandwidth $2W_{l} = 2 (t^{l}_1 + t^{l}_2)$. For simplicity, we choose equal gaps and bandwidths for all leads, $\Delta_{l} = \Delta$ and $W_{l} = W$. In order to keep the chains at half filling, we set their respective chemical potentials to $\mu_{l} = V_{l}$.

Depending on the ratio of the staggered hopping parameters $t^{l}_1$ and $t^{l}_2$, the isolated chain $l$ is either in its normal phase ($t^{l}_1 > t^{l}_2$) or in its topological one ($t^{l}_1 < t^{l}_2$), where it hosts an edge state $\ket{\psi^{l}_{\rm e}}$ with eigenenergy $\epsilon^{l}_{\rm e} = V_{l}$ which is localized near the boundary of the lead \cite{Asboth16}. 
For convenience, we introduce the index $\kappa_{l}$ that labels the phase of lead $l$ as $\kappa_{l} = 1 \,(-1)$ in the topological (normal) phase. 

The term $H_{\rm t} = -\sum_{l} t_{l} a_{l,1}^{\dagger} d \,+\, {\rm H.c.}$ describes the tunnel coupling between the dot and the first site of each lead. We define a characteristic tunneling strength $\Gamma$ considering the gapless metallic limit $t^{l}_1 = t^{l}_2$, where Fermi's golden rule yields $\Gamma = \sum_{l}\Gamma_{l} = \sum_{l} 4 
|t_{l}|^2/W$. This scale shows up as the typical broadening of both the tunneling resonance in transport and the resonance in the dot single-particle spectral function. Note that when we refer to the phase of the SSH lead $l$ in the full, coupled system throughout the article, we always have in mind that of the corresponding isolated chain with $t_l=0$. 

Finally, $H_{\rm int} = U \sum_l  \bigl( a_{l,1}^\dag a_{l,1} - 1/2 \bigr) \bigl( d^\dag  d - 1/2 \bigr)$ accounts for the local two-particle interaction between a fermion on the dot and fermions on the first sites of the leads. Note the shift of the involved density operators by 1/2. It ensures that in equilibrium half-filling implies particle-hole symmetry.

Throughout the article, we consider the regime in which all energy scales are much smaller than the bandwidth $W$, which renders the high-energy details of the leads irrelevant. In this limit $W$ drops out of every $U=0$ calculation. We thus do not give a value for $W$ when presenting our non-interacting results. In contrast, for $U>0$, $W$ matters as a reference scale; see Sect.\ \ref{Sec:Interaction}.

\section{The non-interacting limit \label{Sec:NoInt}}
We begin with the simpler case of vanishing interaction $U = 0$. In this situation, our model can be solved exactly using the Green's function formalism. The details on how to obtain the quantities of interest are discussed in Appendix \ref{App:nonint}, while only the results are presented and analyzed here.

In this section, we identify signatures of the edge states in the dot spectral function, the dot occupation, and in the current across the dot if one or several of the leads are in their respective topological phase. In order to resolve these signatures, we set $\Gamma \ll \Delta$, so that the dot resonance is well defined, i.e., not smeared out, on the scale $\Delta$. Specifically, we set $\Delta = 100 \Gamma$ in following numerical calculations. Different choices lead only to quantitative, but not to qualitative changes, as long as the stated limit is respected.
In Sect.\ \ref{Sec:Interaction}, we consider the robustness of the edge-state signatures against two-particle interaction.

\subsection{The dot spectral function \label{Sec:rho}}

\begin{figure}[t]
\includegraphics[width=0.5\textwidth]{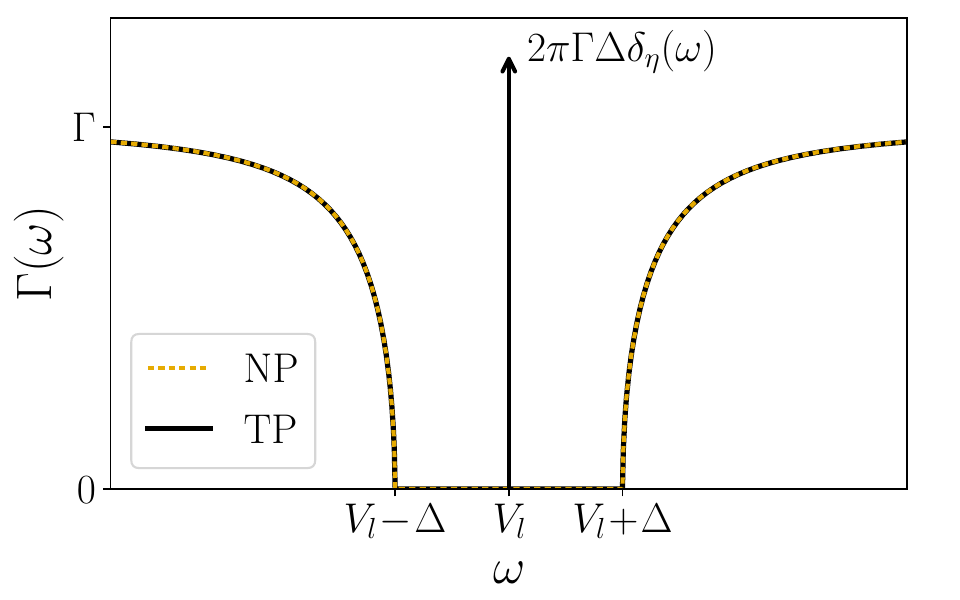}
\caption{The frequency-dependent broadening $\Gamma(\omega)$ of the dot single-particle spectral function induced by a single SSH lead in its normal phase (NP) or its topological phase (TP). It is symmetric about the frequency $\omega = V_l$, where an edge state resides in the topological phase, causing a $\delta$-peak that is symbolized by an arrow in this figure. The broadening vanishes for other frequencies in the gap of the SSH lead, then rises sharply at the two band edges $|\omega - V_l| \geq \Delta$, until it flattens out to a constant $\Gamma$ at large $|\omega-V_l|$.
\label{Fig:Gamma}}
\end{figure}

The dot spectral function is given by
\begin{align}
\rho(\omega) = \frac{1}{\pi} \frac{\Gamma(\omega)/2}{\big[\omega - \epsilon - {\rm Re}\, \Sigma^{\rm R}(\omega)\big]^2 + \Gamma(\omega)^2 / 4} . \label{EQ:rho}
\end{align}
Here $\Sigma^{\rm R}$ denotes the retarded lead self-energy. The spectral function is reminiscent of a Lorentzian; however, the shift of the peak position with respect to $\epsilon$ and the broadening are frequency dependent. Neglecting terms of order $\mathcal{O}(\Delta/W)$, we obtain for the shift
\begin{align}
{\rm Re}\, \Sigma^{\rm R}(\omega) = \sum_{l} \frac{\Gamma_{l}}{2} \left[\kappa_{l}\frac{\Delta_{l}}{\omega_{l}} + \Theta(\Delta_{l} - |\omega_{l}|)\frac{\sqrt{\Delta_{l}^2 - \omega_{l}^2}}{\omega_{l}} \right] , \label{EQ:shift}
\end{align}
with $\omega_{l} = \omega - V_{l}$. The broadening follows from the SSH boundary density of states (isolated lead) $\rho_{l}(\omega)$ via $\Gamma(\omega) = \sum_{l}2 \pi|t_{l}|^2 \rho_{l}(\omega)$ and is given by
\begin{align}
\Gamma(\omega) &= \sum_{l} \Gamma_{l}(\omega) \nonumber \\
& = \sum_{l} \Gamma_{l} \Bigg[ \eta + \Theta(|\omega_{l}| - \Delta_{l})\frac{\sqrt{\omega_{l}^2- \Delta_{l}^2}}{|\omega_{l}|}\nonumber\\
& \hspace{1.6cm} +2\pi\Delta_{l}\delta_{\kappa_{l},1}\delta_{\eta}(\omega_{l})\Bigg]. \label{EQ:broadening}
\end{align}
In Eq.\ (\ref{EQ:broadening}), $\delta_{\eta}(\omega)$ is a nascent $\delta$-function,
\begin{align}
    \delta_{\eta}(\omega) = \frac{1}{\pi}\frac{\eta}{\omega^2 + \eta^2},
    \label{eq:etadelta}
\end{align}
and $\eta >0$ is a small artificial parameter that is taken to zero at the end of any given calculation.
The broadening in a single-lead setup is depicted in Fig.\ \ref{Fig:Gamma}. 

The presence of the gap and of the edge state has several effects on the dot spectral function, which ultimately cause the signatures in the dot occupation and the current that we discuss in subsequent sections.
In the following, we illustrate these effects by means of two examples. For simplicity, we assume that the dot is coupled to a single lead with $V_{l}=0$, and we omit the lead index $l$ in this discussion; a second (or further) lead(s) would simply add to the shift and the broadening in a qualitatively similar fashion. 

\begin{figure}[t]
\includegraphics[width=0.5\textwidth]{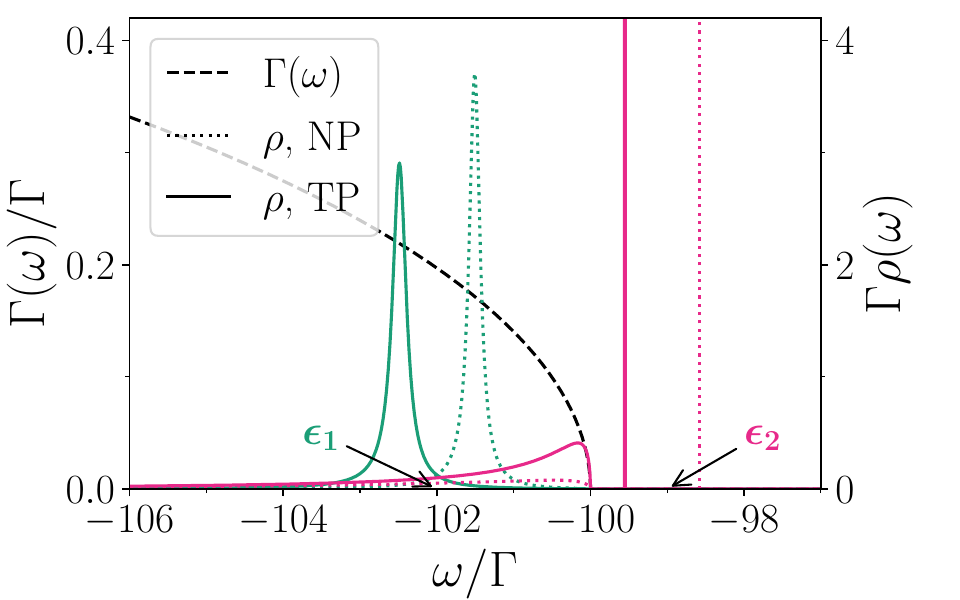}
\caption{The dot single-particle spectral function in the presence of a single lead with $\Delta = 100 \Gamma$ and $V = 0$ (right $y$-axis scale). The dot energy is located slightly below ($\epsilon = \epsilon_1 = -102 \Gamma$, green) or above ($\epsilon = \epsilon_2 = -99\Gamma$, pink; multiplied by a factor of 20 to make the tails visible) the valence band edge. For the dotted (solid) line, the lead is in the normal (topological) phase. The black, dashed curve (left scale) shows the broadening $\Gamma(\omega)$, which is the same for both phases except for a peak at $\omega = 0$ in the topological phase that is not visible in this plot; see Fig.~\ref{Fig:Gamma}. \label{Fig:rho1}}
\end{figure}

Let us start with the situation where the dot energy $\epsilon$ is close to a band edge, see Fig.\ \ref{Fig:rho1}.
When $\epsilon$ lies in the band (green curves in Fig.\ \ref{Fig:rho1}), the dot spectral function consists of a peak which is broadened by the coupling to the continuum. The frequency dependence of $\Gamma(\omega)$  causes an asymmetry of the peak; however, $\Gamma(\omega)$ does not vary much on the scale of the width of the peaks in Fig.\ \ref{Fig:rho1}, so that the asymmetry is weak. Furthermore, there is a shift between the peaks for the two different lead phases (topological/normal), which we will discuss shortly.

For the pink curves in Fig.\ \ref{Fig:rho1}, the dot energy $\epsilon$ lies in the band gap. The resulting peaks are sharp because the broadening $\Gamma(\omega)$ vanishes there, while residual broad tails exist in the band region that vanish $\propto \Gamma(\omega)$ as $\omega$ approaches the band gap from below. As for $\epsilon$ within the band, we see a difference in the peak positions between the two lead phases, which we study in more detail now.

On a technical level, the phase dependence of the peak position can be traced back to the first term in the brackets of the shift Eq.~(\ref{EQ:shift}). To derive an estimate for the splitting in the situation depicted in Fig.\ \ref{Fig:rho1}, we can approximate the width $\Gamma(\omega)$ as constant on the relevant $\omega$ interval around the peak position. This is equivalent to neglecting the peak asymmetry, which is well satisfied for the parameters of Fig.\ \ref{Fig:rho1}. The peak position $\tilde{\epsilon}$ for a given $\epsilon$ then follows from $\tilde{\epsilon} = \epsilon + {\rm Re}\, \Sigma^{\rm R}(\tilde{\epsilon})$. Taking furthermore $\tilde{\epsilon} \approx -\Delta$ in ${\rm Re}\, \Sigma^{\rm R}(\tilde{\epsilon})$ \cite{FN1},
we obtain an estimated peak splitting of $\Gamma$ between the two phases, which fits the one observed in Fig.\ \ref{Fig:rho1} quite well.

We can derive the same estimate from a much more intuitive point of view. We know that the difference between the two lead phases lies chiefly in the presence or absence of the edge state $\ket{\psi_{\rm e}}$, cf. Fig.\ \ref{Fig:Gamma}. This state is coupled to the dot state $\ket{\psi_{\rm d}}$ by the tunneling matrix element $\braket{\psi_{\rm e}|h_{\rm t}|\psi_{\rm d}} = -t\braket{\psi_{\rm e}|1}  = -t\sqrt{4\,\Delta/W}$, where we have labeled by $\ket{1}$ the single-particle state of a fermion on the lead boundary site and inserted the weight of the SSH edge state on the boundary site $\braket{\psi_{\rm e}|1} = \sqrt{4\Delta/W}$. Here $h_{\rm t}$ denotes the `first quantized' (single-particle) version of $H_{\rm t}$. Let us consider an effective model consisting solely of these two states. Diagonalizing the corresponding two-state Hamiltonian
\begin{align}
h = \begin{pmatrix}
\epsilon & -t^* \sqrt{4\Delta / W}\\
-t\sqrt{4\Delta / W} & \epsilon_{\rm e}
\end{pmatrix} \label{EQ:2stateH} .
\end{align}
gives the two eigenstates $\ket{\lambda_{\pm}}$ with eigenenergies
\begin{align}
\lambda_{\pm} = (\epsilon + \epsilon_{\rm e})/2 \pm \sqrt{(\epsilon - \epsilon_{\rm e})^2/4 + \Gamma \Delta} \label{EQ:Eigenenergies}
\end{align}
and weights
\begin{align}
|\braket{\psi_{\rm d}|\lambda_+}|^2 &= (\lambda_+ - \epsilon_{\rm e})^2/\big[(\lambda_+ - \epsilon_{\rm e})^2 + \Gamma \Delta\big] , \label{EQ:weight+}\\
|\braket{\psi_{\rm d}|\lambda_-}|^2 &= \Gamma \Delta/\big[(\lambda_+ - \epsilon_{\rm e})^2 + \Gamma \Delta\big]  \label{EQ:weight-}
\end{align}
on the dot site.

One could now couple these two states to the bands with the properly transformed (change of basis) tunneling parameters, and extract the exact solution for the dot spectral function. For us, however, this is not useful, as we already have the exact solution. The advantage of the two-state model is rather to provide an intuitive picture for the interpretation of the results, which we turn to now. 

For the parameters of Fig.\ \ref{Fig:rho1}, we have $\epsilon_{\rm e} = 0$ and $\epsilon \approx -\Delta$. Thus, the peaks that are visible on the depicted $\omega$ interval correspond to the lower energy $\lambda_-$, which is given by $\lambda_- = \epsilon -\Gamma +\mathcal{O}(\Gamma^2/\Delta)$. That is to say, the dot energy is shifted downwards by roughly $-\Gamma$ by the presence of an edge state. This fits precisely the shift we have found before by analyzing the exact spectral function. Now, however, the origin is much clearer: It is simply the repulsion of the two coupled states $\ket{\psi_{\rm d}}$ and $\ket{\psi_{\rm e}}$ that causes the shift when the lead is in the topological phase.

What about the high-lying state $\ket{\lambda_+}$ that comes out of diagonalizing the two-state Hamiltonian Eq.~(\ref{EQ:2stateH})? 
Its energy is $\lambda_+ = \epsilon_{\rm e} + \Gamma + \mathcal{O}(\Gamma^2/\Delta) \approx \Gamma$, so in Fig.~\ref{Fig:rho1} it lies outside the 
shown $\omega$ window. In any case, it has a negligible weight on the dot site, $|\braket{\psi_{\rm d}|\lambda_+}|^2 = \Gamma/\Delta + \mathcal{O}(\Gamma^2/\Delta^2) \ll 1$.

\begin{figure}[t]
\includegraphics[width=0.5\textwidth]{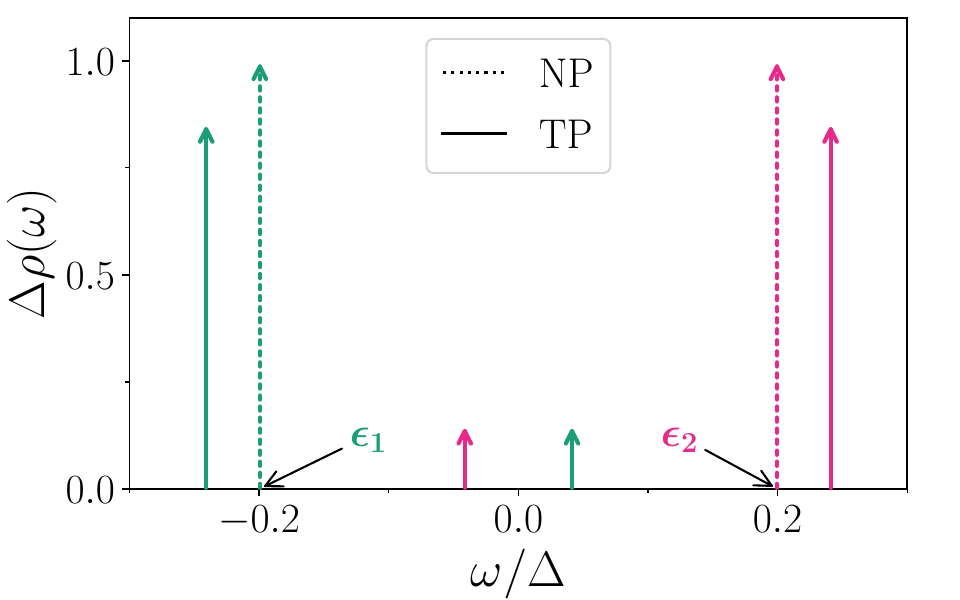}
\caption{The dot single-particle spectral function in the presence of a single lead with $\Delta = 100 \Gamma$ in the normal (dotted lines) or topological phase (solid lines). The dot energy is inside the gap, below ($\epsilon = \epsilon_1 = -0.2 \Delta$, green) or above ($\epsilon = \epsilon_2 =0.2 \Delta$, pink) the edge state energy $\epsilon_{\rm e} = 0$. As the broadening vanishes here, the peaks are perfectly sharp $\delta$-functions. The height of the arrows represents their weights.}
\label{Fig:rho2}
\end{figure}

For sufficiently small energy differences $\epsilon - \epsilon_{\rm e} \sim \sqrt{\Gamma \Delta}$, though, the two-state calculation suggests that the second state does have a sizeable weight on the dot site, which should translate to an appreciable second peak in the dot spectral function. This is indeed the case. In Fig.~\ref{Fig:rho2} the exact dot spectral function is shown for two dot energies that lie rather close to $\epsilon_{\rm e} = 0$. As the peaks fall in frequency regions with vanishing broadening $\Gamma(\omega)$, they are $\delta$-peaks; the height of the arrows in Fig.\ \ref{Fig:rho2} symbolizes the weight of the peaks. As a result of the hybridization of the dot and the edge state, we indeed observe a two-peak structure in the exact dot spectral function when the lead is in the topological phase, just as the two-state picture suggests. 

The emergence of the second peak can also be shown formally, starting from the exact formula for the dot spectral function Eq.~(\ref{EQ:rho}). Let us again take $\Gamma(\omega) = \rm const.$. Technically, this suppresses the peak asymmetry for peaks lying in the band region, but we are only interested in the existence of peaks now, not their detailed line shape. For frequencies in the gap, $\Gamma(\omega) = 0^+  = \rm const.$ is given everywhere except for $\omega = V = 0$, where $\Gamma(\omega)$ becomes very large. However, we can safely exclude $\omega=0$ in the following analysis, as no peak of $\rho(\omega)$ can lie there. This is because the very large $\Gamma(\omega = 0)$ appears linearly in the numerator and quadratically in the denominator, compare Eq.~(\ref{EQ:rho}).

With the assumption of constant broadening, the equation for the peak position $\tilde{\epsilon}$ of $\rho(\omega)$ is $\tilde{\epsilon} = \epsilon + {\rm Re}\, \Sigma^{\rm R}(\tilde{\epsilon})$. With the lead in the normal phase, $|{\rm Re} \,\Sigma^{\rm R}(\omega)|$ is bounded by $\Gamma/2$ and is piecewise strictly monotonic; its derivative changes sign at $\pm \Delta$. This leads to a single solution for the peak position at $\tilde{\epsilon} = \epsilon + \mathcal{O}(\Gamma)$ \cite{FN2}. 
In contrast, with the lead in the topological phase, ${\rm Re} \,\Sigma^{\rm R}(\omega)$ has a $\Gamma \Delta / \omega$ divergence as $\omega \rightarrow 0$, which leads to an additional solution for $\tilde{\epsilon}$ close to zero (or in general close to $V$--but never at exactly $\tilde{\epsilon} = V =0$, cf. the discussion above) \cite{FN3}.
Finally, we note that this reasoning applies directly to any further edge states if multiple leads are attached to the dot: For every topological lead $l$, there is a $\Gamma_{l} \Delta_{l} / \omega_{l}$ divergence in ${\rm Re} \,\Sigma^{\rm R}(\omega)$, leading to an additional peak solution close to $V_{l}$.

We conclude that a simple two-state picture of the dot and the edge state, which hybridize and repel each other, is a very useful effective model to understand the dominant signatures of the edge state in the dot spectral function: the shift of the dot resonance and the emergence of a second peak close to the edge-state energy. Essentially, the edge state acts as an additional side-coupled dot. These signatures of the edge state have an impact on the dot occupation in equilibrium, which we discuss in the next section, as well as on the current across the dot when multiple leads are attached and a bias voltage is applied, cf.\ Sect.~\ref{Sec:current}.

\subsection{The dot occupation \label{Sec:occupation}}

\begin{figure}[t]
\includegraphics[width=0.5\textwidth]{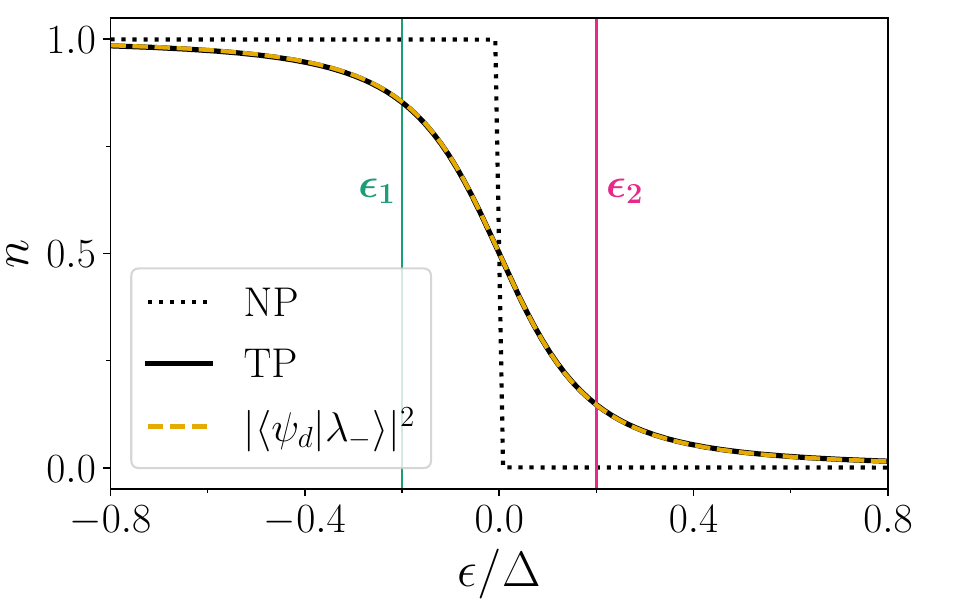}
\caption{The dot occupation as a function of the level position in the presence of a single SSH lead with $\Delta = 100 \Gamma$ at $\mu = 0$. 
The results of the exact calculations are shown as black dotted (normal phase) and solid (topological phase) lines; the dashed orange line corresponds to the prediction of the two-state model with the lead in the topological phase.
The two energies $\epsilon_1$ and $\epsilon_2$ marked by vertical lines are those for which the dot spectral function is shown in Fig.~\ref{Fig:rho2}.}
\label{Fig:occupation}
\end{figure}

Without loss of generality, we study again a single SSH lead at zero chemical potential, i.e., $\mu = V = 0$, coupled to the dot. The dot occupation $n$ is then given by
\begin{align}
n = \int_{-\infty}^0 d\omega \, \rho(\omega). \label{EQ:occupation}
\end{align}
The occupation as a function of the dot energy $\epsilon$ is shown in Fig.\ \ref{Fig:occupation}. It changes from one to zero as $\epsilon$ crosses the chemical potential of the lead, i.e., around $\epsilon = 0$.

Consider first the situation of an SSH chain in its normal phase. For the dot energies shown in Fig.\ \ref{Fig:occupation}, the dot spectral function consists of a single sharp peak within the gap, cf. Fig.\ \ref{Fig:rho2}, while the tails in the band regions can be neglected, as $\Gamma \ll \Delta$. Crucially, the entire weight of $\rho(\omega)$ lies either below (when $\epsilon<0$) or above (when $\epsilon>0$) the lead chemical potential. Therefore, only the values $n = 1$ and $n = 0$ are possible, and a sharp decharging transition occurs between the two at the particle-hole symmetric point.

When the lead is topological, the two-peak structure of $\rho(\omega)$ discussed in Sect.~\ref{Sec:rho} changes the situation drastically. As $\epsilon$ approaches the chemical potential $\mu$ from below, it simultaneously comes closer to the edge-state energy $\epsilon_{\rm e} = \mu = 0$. This causes the hybridization of the dot and the edge state to increase, which is accompanied by an increasing peak in $\rho(\omega)$ above $\epsilon_{\rm e}$, and thus above $\mu$. Due to the sum rule $\int d\omega \, \rho(\omega) = 1$, the weight of the peak below $\mu$ must decrease, which causes the occupation Eq.~(\ref{EQ:occupation}) to decrease continuously. Similarly, a peak in $\rho(\omega)$ below the chemical potential remains even when $\epsilon > \mu = 0$, so that $n(\epsilon)$ drops smoothly to zero as the weight of this peak decreases.

To illustrate this further, let us consider the effective two-state model calculation of Sect.~\ref{Sec:rho}, which gave a state $\ket{\lambda_+}$ above and a state $\ket{\lambda_-}$ below the chemical potential $\mu = \epsilon_{\rm e}$. The resulting dot occupation is $n(\epsilon) = |\braket{\psi_{\rm d}|\lambda_-}|^2$, given in Eq.\ (\ref{EQ:weight-}). This is shown in Fig.\ \ref{Fig:occupation} as the orange dashed line. It almost perfectly fits the exact result. This reinforces our insight that the principal mechanism causing the broadening is the hybridization of the dot and the edge state, which is accounted for in a simple two-state picture. Furthermore, the results of the two state-calculation, Eqs.\ (\ref{EQ:weight-}) and (\ref{EQ:Eigenenergies}), reveal that the scale of the broadening is given by $\sqrt{\Gamma \Delta}$.

We stress that the broadening of the dot occupation when varying the onsite energy results from the hybridization of the two states, and not from the coupling of the dot level to a continuum of states such as the leads' energy bands.

\subsection{The current \label{Sec:current}}

Another strong signature of the edge state is found in the finite-bias current across the dot. To study this, we consider a setup with two leads [left (L) and right (R)] which are coupled asymmetrically ($\Gamma_{\rm L} = 3\Gamma/5$, $\Gamma_{\rm R} = 2\Gamma/5$) to the dot so as to represent a generic coupling situation one might encounter in experiments. A bias voltage $V_{\rm L} = -V_{\rm R} = V/2$ is applied. The steady-state current $I$ from left to right across the dot is (see Appendix \ref{App:nonint}) 
\begin{align}
I = \int_{-V/2}^{V/2} d\omega \,\frac{\Gamma_{\rm L}(\omega)\Gamma_{\rm R}(\omega)}{\Gamma(\omega)}\rho(\omega) . \label{EQ:current}
\end{align}

The factor $\Gamma_{\rm L}(\omega)\Gamma_{\rm R}(\omega)$ in Eq.~(\ref{EQ:current}) implies that a current can flow only when the left valence band (with energies below $V_{\rm L}-\Delta)$ overlaps with the right conduction band (with energies above $V_{\rm R} + \Delta$). This happens at voltages $V/ 2 > \Delta$. 
As $\rho(\omega)$ appears in Eq.~(\ref{EQ:current}), the current increases significantly when the broadened peak in the dot spectral function lying close to $\epsilon$ enters this region of overlap. This occurs at voltages around $V/2 = \Delta + |\tilde{\epsilon}|$, with the peak position $\tilde{\epsilon}$. Note that any further peak induced by possible edge states (in the leads) lies close to $V_{\rm L}$ or $V_{\rm R}$, as we have found in Sect.~\ref{Sec:rho}, and thus can never enter the overlap region. 

\begin{figure}[t]
\includegraphics[width=0.5\textwidth]{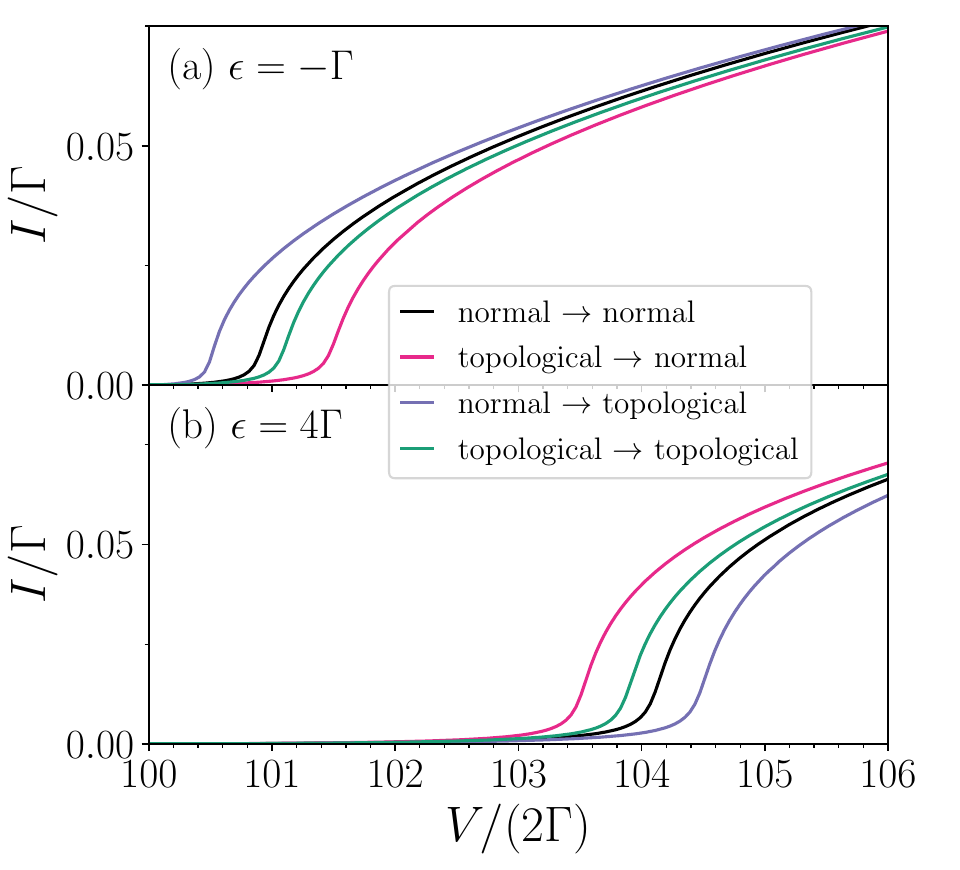}
\caption{The current as a function of the bias voltage in a two-terminal setup with $V_{\rm L} = -V_{\rm R} = V/2$, $\Gamma_{\rm L} = 3\Gamma/5$, $\Gamma_{\rm R} = 2\Gamma/5$, and $\Delta = 100 \Gamma$, for (a) a negative or (b) a positive dot energy and varying lead phases, encoded by the line colors. The arrows in the legend represent the current direction (from the left lead to the right lead). For large voltages (not shown here), the current levels off to a constant.}
\label{Fig:current}
\end{figure}

Figure \ref{Fig:current} shows the current for various lead configurations and two different dot energies. Let us take the case with two normal leads as a baseline (black lines in Fig.\ \ref{Fig:current}) and denote by $\tilde{\epsilon}_0$ the corresponding peak position in the dot spectral function. We find that the current onset is shifted relative to the one in this normal case if any lead is topological. This is easily understood in terms of the shift of the peak in $\rho(\omega)$ discussed in Sect.~\ref{Sec:rho} and exemplified in Fig.~\ref{Fig:rho1}. 

Consider the situation where the left lead is topological while the right lead is normal (pink lines in Fig.~\ref{Fig:current}). To understand the difference between this case and the baseline established in the previous paragraph, we have to examine the effect of the edge state that now exists in the left lead. The edge state pushes the peak position $\tilde{\epsilon}$ downwards from $\tilde{\epsilon}_0$ by roughly $-\Gamma_{\rm L} = -3\Gamma/5$. The resulting shift in the current onset depends on $\tilde{\epsilon}_0$. If $\tilde{\epsilon}_0$ is negative, then $|\tilde{\epsilon}|$ is increased by the edge state, causing the current onset at $V/2 = \Delta + |\tilde{\epsilon}|$ to occur at higher voltages. If $\tilde{\epsilon}_0$ is positive (and greater than approximately $\Gamma_{\rm L}/2$, to be precise), $|\tilde{\epsilon}|$ is reduced by the edge state, which leads to a current onset at lower voltages. The dot energies in Fig.~\ref{Fig:current} are such that they reflect these two possibilities.

If, on the other hand, the left lead is normal while the right lead is topological (blue lines in Fig.\ \ref{Fig:current}), then the right edge state pushes the peak position $\tilde{\epsilon}$ upwards from $\tilde{\epsilon}_0$ by roughly $\Gamma_{\rm R} = 2\Gamma/5$, as the edge state lies at a lower energy than the dot state in this configuration. As a consequence, the effect is reversed with respect to the previously discussed configuration, and the magnitude is different because $\Gamma_{\rm L} \neq \Gamma_{\rm R}$.

Finally, if both leads are topological (green lines in Fig.\ \ref{Fig:current}), the two edge states push the peak in $\rho(\omega)$ in opposite directions. We do not have to carry out the calculation of Sect.\ \ref{Sec:rho} with three states---the dot and the two edge states---to see this. Instead, we can employ perturbation theory in the tunnel coupling and find that each edge state $\ket{\psi_{\rm e}^{l}}$ shifts the dot energy by roughly $(\epsilon - \epsilon_{\rm e}^{l})^{-1} \, |t_{l}|^2 \, 4\Delta/W  = \Gamma_{l} \Delta/(\epsilon-V_{l})$ \cite{FN4}.
Regarding the voltage setup and dot energies considered in Fig.\ \ref{Fig:current}, we can estimate for voltages around the current onset that $\epsilon-V_l \approx \mp\Delta$ for $l=\rm{L,R}$, which leads to an edge-states-induced shift of the peak position $\tilde{\epsilon}$ of approximately $\Gamma_{\rm R}- \Gamma_{\rm L}$. With our choice of couplings $\Gamma_{\rm L} > \Gamma_{\rm R}$, this shift is negative, leading to an anticipated current onset if $\tilde{\epsilon}_0 > 0$, and a retarded onset if $\tilde{\epsilon}_0 < 0$. The opposite would happen if $\Gamma_{\rm L} < \Gamma_{\rm R}$, while the shifts would approximately cancel out if one chose equal couplings $\Gamma_{\rm_L} = \Gamma_{\rm R}$.

In conclusion, with topological leads, the current can have a retarded, early, or unaltered onset as compared to the case with normal leads, depending on the precise configuration of edge states. All cases can be explained and understood by considering the repulsion of the dot and the edge states.

\section{Including the two-particle interaction \label{Sec:Interaction}}

We now investigate if the effects of the topological edge states on the dot spectral function, the dot occupation, and the finite-bias current across the dot are robust against a local two-particle interaction $U$. This will turn out to be the case. In fact, the effects on the spectral function and the current will even be enhanced. 

To this end we compute the interacting part of the self-energy, $\Sigma^{\rm int}$, in first-order perturbation theory in $U$, employing either Matsubara (equilibrium) or Keldysh (non-equilibrium) formalism. 
Because of the locality of the interaction, only a few matrix elements of the self-energy are non-vanishing. In the notation used in Sect.~\ref{sec:model} for the lead lattice sites and denoting the dot site by the label d, these are $\Sigma^{\rm int}_{{\rm d},{\rm d}}$, $\Sigma^{\rm int}_{(l,1), (l,1)}$, and $\Sigma^{\rm int}_{{\rm d}, (l,1)}$. 

To leading order, $\Sigma^{\rm int}$ is frequency independent. Thus, the matrix elements can be interpreted as interaction-induced changes of the corresponding single-particle parameters of the non-interacting Hamiltonian. In this sense, $\Sigma^{\rm int}_{{\rm d},{\rm d}}$ and $\Sigma^{\rm int}_{(l,1), (l,1)}$ are the interaction-induced changes of the onsite energies of the dot site (non-interacting value $\epsilon$) and of the first lead lattice site (non-interacting value $V_l$), respectively. They follow from the Hartree diagram. The third non-vanishing matrix element $\Sigma^{\rm int}_{{\rm d}, (l,1)}$ captures the change of the tunnel coupling (non-interacting value $t_l$) and results from the Fock diagram.

It is straightforward to obtain numerical results for $\Sigma^{\rm int}$ and, based on these, calculate the observables of interest to us. In contrast to the non-interacting case, $W$ enters as a reference scale when $U>0$, even if the band width is taken much larger than any other energy scale in the problem. Details on the calculations and a brief discussion on the explicit appearance of $W$ can be found in Appendix \ref{App:1O}.

In the following, we present and discuss numerical results for the observables in first-order perturbation theory. Because of the aforementioned $W$ dependence, we shall always specify the value of $W$ considered. Values of $U$ are chosen such that the modifications due to the interaction are visible on the scales of the plots. For a brief discussion on the question if, for these $U$, first-order perturbation theory for the self-energy can be trusted quantitatively, we refer to Sec.\ \ref{Sec:Conclusion}.
Besides, we refrain from tracing back the interaction dependence of the observables to the formulas presented in Appendix \ref{App:1O} analytically, though it is possible in principle. For the robustness check we are aiming at, a qualitative understanding suffices.

\subsection{The dot spectral function}
\label{subsec:eff_dsf}

\begin{figure}[t]
\includegraphics[width=0.5\textwidth]{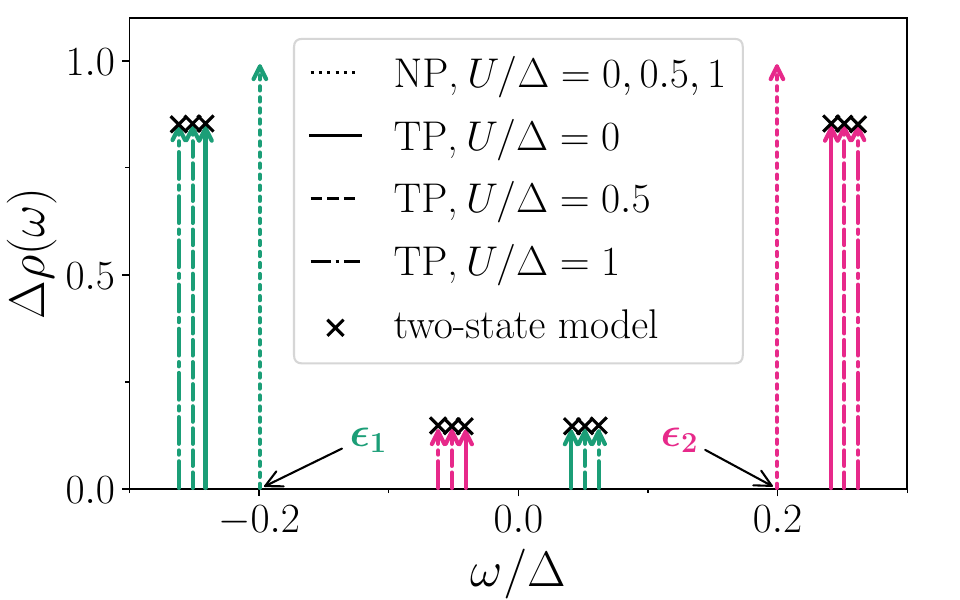}
\caption{The dot spectral function in the presence of a single lead with $\Delta = 100 \Gamma = 0.01W$ for two different dot energies $\epsilon = \epsilon_1 =-0.2\Delta$ (green) and $\epsilon = \epsilon_2 =0.2\Delta$ (pink), analogous to Fig.\ \ref{Fig:rho2}. Different interaction strengths $U/\Delta=0,0.5,1$ are considered. As the peaks lie in the energy gap of the lead, they are perfectly sharp $\delta$-functions; we represent their weight as the height of the arrows.
The data calculated from the two-state model are shown for comparison (black crosses). }
\label{Fig:rho_int}
\end{figure}

In discussing the effect of the interaction on the dot spectral function, we limit ourselves to the situation shown in Fig.\ \ref{Fig:rho2}, where the dot and the edge-state energy are close to one another, and the dot is coupled to a single lead with $\mu = V = 0$. In this scenario both the two-peak structure due to the hybridization as well as the shift of the dot level peak are visible. We do not discuss the situation with the dot energy close to the band edge, depicted in Fig.\ \ref{Fig:rho1}, as the interaction effect can be seen directly in the current in that case, cf.\ Sect.\ \ref{sec:interactingcurrent}. 
Figure \ref{Fig:rho_int} shows the spectral function with the same parameters as Fig.\ \ref{Fig:rho2}, but with several values for the interaction strength.

Let us start with the SSH lead being in its normal phase (dotted arrows). For this the peak is always located at $\epsilon$ and the weight is $U$-independent as well; the arrows for different $U$ fully overlap and cannot be distinguished. 

In contrast, when the lead is in its topological phase, the shift of the peaks is enhanced linearly in $U$ for $U > 0$, while the distribution of the spectral weight remains independent of $U$ (to leading order). This holds for both $\epsilon<0$ (green arrows) and $\epsilon>0$ (pink arrows).

As in the non-interacting case, the effective two-state picture, with the bare parameters replaced by the ones complemented by the self-energy, can be used as well; see Appendix \ref{App:1O} for details. As Fig.~\ref{Fig:rho_int} illustrates, the peak positions as well as the corresponding weights obtained from this model (black crosses) agree nicely with the ones derived from the full calculation (arrows), which shows that the two-state model remains a useful intuitive picture in the presence of interaction.

\subsection{The dot occupation}
\label{subsec:eff_Dot_Occupation}

Regarding the effect of a two-particle interaction on the dot occupation to leading order in $U$, we can be very brief: There is none. For the case of a lead in its normal phase this is obvious from the spectral function discussed in the last section. The peak position and height is independent of $U$, which by Eq.~(\ref{EQ:occupation}) (which also holds for $U>0$) directly translates into an occupation that is interaction independent. If the lead is topological and features an edge state, the interaction effect is negligible as well. This is because only the weight of the spectral function below $\omega=0$ matters for the dot occupation; cf.\ Eq.\ (\ref{EQ:occupation}). This is unchanged by the interaction; compare Fig.~\ref{Fig:rho_int}. Note that the perfect match between the two-state model calculation and the full result for the spectral function implies that both give the same dot occupation as well.

\subsection{The current}
\label{sec:interactingcurrent}

\begin{figure}[t]
\includegraphics[width=0.5\textwidth]{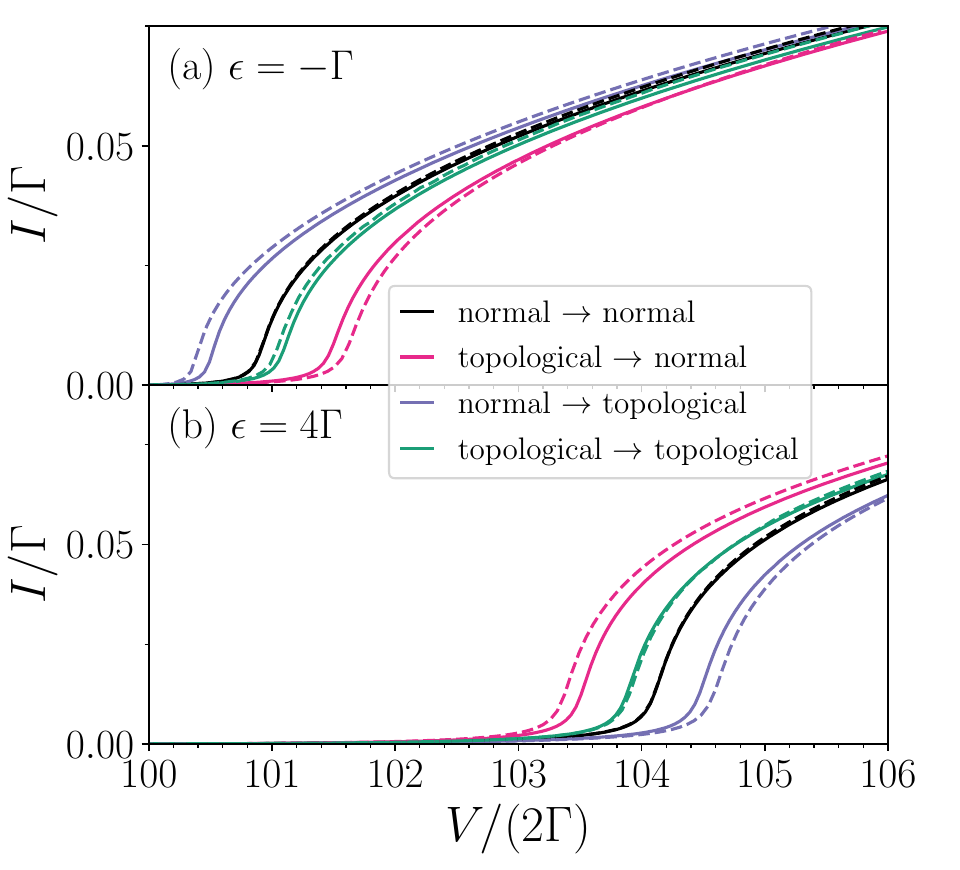}
\caption{The current as a function of the bias voltage for voltages of the order of the gap, with $V_{\rm L} = -V_{\rm R} = V/2$, $\Gamma_{\rm L} = 3\Gamma/5$, $\Gamma_{\rm R} = 2\Gamma/5$, and $\Delta = 100 \Gamma = 4.5\times10^{-4}W$, for two different single-particle dot energies, analogous to Fig.\ \ref{Fig:current}. The interaction strength is $U=0$ (solid lines) and $U=160\Gamma$ (dashed lines).}
\label{Fig:current_frg}
\end{figure}

We finally study the bias-voltage-driven non-equilibrium current in a two-lead setup. For the corresponding relation between Green's functions and the current, see Appendix \ref{App:1O}. Note that the voltage does not only enter via the boundaries of the integration as in Eq.~(\ref{EQ:current}), but additionally via the self-energy, which picks up a $V$-dependence due to the interaction.

Figure~\ref{Fig:current_frg} shows the current for the same single-particle parameters as in Fig.~\ref{Fig:current} but with the interaction strength $U=160\Gamma$ in addition to the $U=0$ results. The $U=0$ data are shown as solid lines, while the ones for $U>0$ are shown as dashed lines. Similar to what we have found for the dot spectral function, the interaction enhances the effect caused by the hybridization between the dot and the edge state: When only one of the SSH leads is in its topological phase, the current onset is shifted even further compared to the non-interacting case. We also find a slight modification of the non-interacting result in the case of two topological leads. This modification is largely due to the asymmetric couplings, it would be reduced in the symmetric case $\Gamma_{\rm L} = \Gamma_{\rm R}$. At increased voltages, all curves approach each other.

\section{Conclusion}
\label{Sec:Conclusion}

We have investigated a model consisting of a single-level quantum dot weakly tunnel coupled to one or two fermionic SSH leads, focusing on signatures of topological edge state(s) when the leads are in their respective topological phase. Starting with the dot single-particle spectral function, we have found two signatures of the edge state(s): a shift of the peak associated with the dot excitation and the emergence of a second peak close to the edge-state energy.
Employing the effective two-state model of the dot and the edge state, it is straightforward to see that both these effects stem from the hybridization of these two states. 

Furthermore, we have identified consequences of this in the dot occupation and in the electric current. 
The emergence of the second peak strongly broadens the step in the dot occupation as a function of the dot energy, which is sharp when the lead is topologically trivial.
The shift of the dot peak, on the other hand, can be measured in the current across the dot, where, depending on the parameters, it leads to a retardation or an early onset of the current when varying the bias voltage.

In the second part of the paper, we have studied whether these effects are robust against a local two-particle interaction in the dot region. Our results for the dot single-particle spectral function, the dot occupation, and the small-bias-voltage current show that all edge-state signatures in these observables are, to leading order in $U$, robust or even enhanced. 
Furthermore, the fact that the full spectral function agrees with the prediction of the two-state model demonstrates that the latter remains a valid notion, provided that one inserts parameters which take into account the interaction-modified self-energy.

Note that our leading-order calculation only serves to ensure that the effects identified at vanishing interaction do not break down at weak interaction. In order to investigate for which values of system parameters first-order perturbation theory can be trusted quantitatively, one would have to compare our results to those obtained by employing more sophisticated analytical many-body methods that include higher orders of $U$, or to those of numerical methods. 
Using such methods would also allow studying non-perturbative correlation effects. For example, one might expect correlation effects found in the interacting resonant level model (see, e.g., \cite{Karrasch10}) to carry over to the setup studied here, as $\Delta$ cuts off the renormalization flow  similar to other model parameters like $\epsilon$ or $V$ [see Eq.\ (\ref{eq:hoprenex})]\cite{Lin22}.

The edge-state signatures in the current and the dot occupation can be used to determine the presence of an edge state and thus to determine the phase of the SSH lead(s). 
Regarding the current, the $I$-$V$ curve looks qualitatively the same between the two phases. The difference is a relative shift of the curve and, therefore, well visible only when the phase is switched between measurements.
In contrast, a single measurement of the dot occupation suffices to determine the phase: A sharp step corresponds to a normal lead, a broadened step to a topological one. Therefore, the occupation--measurable via the current flowing through a quantum point contact close to the dot region \cite{Hanson07}--constitutes a simple experimental probe and an alternative to the method presented in Ref.\ \onlinecite{Zaimi21}, where the decoherence of a double dot has been proposed as an edge-state detector.
We note that spectral and transport characteristics of quantum dots were also used in the context of Majorana bound state detection \cite{Liu11,Vernek14,Ruiz-Tijerina15,Deng16,Hoffman16,Hoffman17}.

We emphasize that we do not expect the discussed edge-state effects to be specific to the topological nature of the edge states. However, it would be interesting to investigate this in the future by considering, for example, Rice-Mele model leads \cite{Rice82} instead of SSH ones, in which edge states are not directly related to topological properties. Furthermore, it would be interesting to study a dot with multiple levels, e.g., including the spin degree of freedom, and to investigate how a local interaction on the dot influences the effects we have found in this article.

\begin{acknowledgments}
This  work  was  supported by the Deutsche Forschungsgemeinschaft (DFG, German  Research  Foundation)  via RTG  1995. DMK acknowledges support from the Max Planck-New York City Center for Non-Equilibrium Quantum Phenomena.
The numerical calculations have been performed with computing resources granted by RWTH Aachen University under project rwth0444.
\end{acknowledgments}

\appendix

\section{Details on the non-interacting case \label{App:nonint}}

Here we give a brief summary of the relevant quantities for our non-interacting model. In this case all observables of interest to us can be expressed in terms of the dot Green's function.
It is defined as
\begin{align}
G^{{\rm R/A}}(t) &= \mp i \Theta(\pm t) \langle\lbrace d(t), d^{\dagger} \rbrace\rangle \\
&= \frac{1}{2\pi}\int d\omega \,e^{-i\omega t} G^{{\rm R/A}}(\omega).
\end{align}
The Fourier transform can be written in the usual way in terms of a lead self-energy $\Sigma (\omega)$,
\begin{align}
G^{{\rm R/A}}(\omega) = \frac{1}{\omega - \epsilon - \Sigma^{{\rm R/A}}(\omega)} .
\end{align}
The lead self-energy can be expressed as
\begin{align}
\Sigma^{{\rm R/A}}(\omega) &= \sum_l \Sigma_l^{\rm R/A}(\omega) = \sum_{l} \braket{\psi_{\rm d}|h_{\rm t}^{l}\, g^{{\rm R/A}}_{l}(\omega)\, h_{\rm t}^{l}|\psi_{\rm d}} \nonumber \\
&= \sum_{l}|t_{l}|^2 \braket{1_{l}|g^{{\rm R/A}}_{l}(\omega)|1_{l}},
\label{eq:leadselfenergy}
\end{align}
with the free retarded/advanced lead Green's function
\begin{align}
g^{{\rm R/A}}_{l}(\omega) =\frac{1}{\omega - h_{l} \pm i\eta}.
\end{align}
Here $h_{\rm t}^l$ and $h_l$ denote the `first quantized' (single-particle) versions of $H_{\rm t}^l$ and $H_l$, respectively, and the convergence factor $\eta > 0$ is taken to zero at the end of any calculation.

The lead boundary Green's function can be computed as follows. In matrix representation the single-particle Hamiltonian of a single SSH lead (w.l.o.g.\ $V_l = 0$, and the index $l$ is suppressed) is given as
\begin{align} \label{eq:semi_inf_ssh}
h =
\left( \begin{array}{cccc}
0 & -t_1 &    & \phantom{-}  \\[8pt]
-t_1 & 0 & -t_2 & \phantom{--}  \\
\phantom{--} & -t_2& 0 & \ddots  \\
\phantom{--} &  \phantom{--} & \ddots & \ddots  \\
\phantom{--} &  \phantom{--} & \phantom{--} & \phantom{--} 
\end{array}\right) .
\end{align}
Due to the periodic tridiagonal structure of this semi-infinite matrix, one can find the top left entry of the inverse analytically to obtain the boundary Green's function,
\begin{align}
\braket{1|g^{\rm R/A}(\omega)|1}&=\left( \frac{1}{z-h} \right)_{11} \nonumber\\ &=\frac{1}{2z\, t_2^2} \Bigl\{ z^2-t_1^2+t_2^2 \nonumber \\
&\hspace{1.5em} - \sqrt{z+W}\sqrt{z-W}\sqrt{z+\Delta}\sqrt{z-\Delta} \Bigr\}, \label{eq:boundary_gf}
\end{align}
with $z=\omega\pm i\eta$ for the retarded/advanced Green's function.
In the limit $\omega,\Delta \ll W$, we can write 
\begin{align}
&\text{Re} \braket{1|g^{\rm R}(\omega)|1}=\frac{2}{W} \left[\kappa\frac{\Delta}{\omega} + \Theta(\Delta - |\omega|)\frac{\sqrt{\Delta^2 - \omega^2}}{\omega} \right], \nonumber
\\ 
&\text{Im} \braket{1|g^{\rm R}(\omega)|1}=-\frac{2}{W}\Bigg[\eta +  \Theta(|\omega| - \Delta)\frac{\sqrt{\omega^2- \Delta^2}}{|\omega|}\nonumber \\ & \hspace{3.3cm} + 2\pi \Delta\delta_{\kappa,1}\delta_{\eta}(\omega)\Bigg], 
 \label{eq:boundary_gf_scaling_limit}
\end{align}
where $\kappa=1(-1)$ labels the topological (normal) phase. The broadened $\delta$-function $\delta_{\eta}(\omega)$ is defined in Eq.~(\ref{eq:etadelta}). The advanced Green's function is obtained by complex conjugation: $g^{\rm A}(\omega) = g^{\rm R}(\omega)^*$.

When the chain is in its topological phase, the boundary Green's function has a pole at $\omega = 0$ corresponding to the zero-energy edge state,
\begin{align}
\braket{1|g^{\rm R}(\omega)|1} \xrightarrow[]{\omega \rightarrow 0} |\langle 1 | \psi_{\text{\rm e}} \rangle|^2 \frac{1}{\omega + i \eta},
\end{align} 
with the weight $\braket{ 1 |\psi_{\text{\rm e}} } = \sqrt{4\Delta/W}$ on the boundary site.

From the knowledge of the retarded dot Green's function we can calculate the dot spectral function
\begin{align}
\rho(\omega) = -\frac{1}{\pi}{\rm Im}\, G^{\rm R}(\omega).
\label{EQ:dot_spectral_function}
\end{align}
Decomposing the retarded lead self-energy into a real and imaginary part,
\begin{align}
\Sigma^{\rm R}(\omega) = {\rm Re}\, \Sigma^{\rm R}(\omega) - i \Gamma(\omega)/2,
\end{align}
we obtain Eqs.~(\ref{EQ:rho})--(\ref{EQ:broadening}).

The steady-state current flowing from left to right in a two-terminal setup is given by \cite{MW92}
\begin{align}
I =\int d\omega \, T(\omega)\big[f_{\rm L}(\omega)-f_{\rm R}(\omega) \big] , \label{EQ:AppCurrent}
\end{align}
with the transmission coefficient
\begin{align} \label{EQ:transmission_coefficient}
T(\omega) &= \frac{1}{2\pi}\Gamma_{\rm L}(\omega)G^{\rm A}(\omega)\Gamma_{\rm R}(\omega)G^{\rm R}(\omega) \nonumber \\
&= \frac{\Gamma_{\rm L}(\omega)\Gamma_{\rm R}(\omega)}{2\pi} |G^{\rm R}(\omega)|^2 \nonumber \\
&= \frac{\Gamma_{\rm L}(\omega)\Gamma_{\rm R}(\omega)}{2\pi} \frac{1}{\big(\omega - \epsilon - {\rm Re}\, \Sigma^{\rm R}(\omega)\big)^2 + \Gamma(\omega)^2 / 4} \nonumber\\
&= \frac{\Gamma_{\rm L}(\omega)\Gamma_{\rm R}(\omega)}{\Gamma(\omega)} \left[\frac{1}{\pi} \frac{\Gamma(\omega)/2}{\big(\omega - \epsilon - {\rm Re}\, \Sigma^{\rm R}(\omega)\big)^2 + \Gamma(\omega)^2 / 4}\right] \nonumber \\
&= \frac{\Gamma_{\rm L}(\omega)\Gamma_{\rm R}(\omega)}{\Gamma(\omega)} \rho(\omega) .
\end{align}
Inserting this into (\ref{EQ:AppCurrent}) together with $f_{\rm L}(\omega) - f_{\rm R}(\omega) = \Theta(V/2 - \omega) - \Theta(-V/2 - \omega)$ at zero temperature and symmetric bias, we obtain Eq.~(\ref{EQ:current}) for the current.

\section{First-order perturbation theory\label{App:1O}}

Here we present the basic equations to compute the interacting part of the self-energy $\Sigma^{\rm int}$ to first-order perturbation theory in $U$ and, from this, the observables of interest to us. Equilibrium as well as the bias-voltage-driven non-equilibrium steady state are considered.

\subsection{Equilibrium}

We first consider the setup of a single lead coupled to the dot level in equilibrium. Suppressing the lead index $l$, the three non-vanishing and frequency independent matrix elements of $\Sigma^{\rm int}$ in Matsubara formalism read 
\begin{align}\label{EQ:PT_Matsubara_dd}
    &\Sigma^{\text{int}}_{\text{d,d}}=-\frac{U}{\pi}\int^{\infty}_0 d\omega \, G^{\text{eq}}_{\text{1,1}}(i\omega) ,\\ \label{EQ:PT_Matsubara_d(1l)}
    &\Sigma^{\text{int}}_{1,1}=-\frac{U}{\pi}\int^{\infty}_0 d\omega \, G^{\text{eq}}_{\rm d,d}(i\omega), \\
    &\Sigma^{\text{int}}_{\text{d},1}=\frac{U}{\pi}\int^{\infty}_0 d\omega \, G^{\text{eq}}_{{\rm d},1}(i\omega).
    \label{EQ:PT_Matsubara_(1l)(1l)}
\end{align}
Here $\Sigma^{\text{int}}_{\text{d,d}}$ and $\Sigma^{\text{int}}_{1,1}$ are the change of the dot-level energy (non-interacting value $\epsilon$) and the first lead lattice site (non-interacting value $V=0$), respectively. The off-diagonal matrix element $\Sigma^{\text{int}}_{\text{d},1}$ is the  interaction correction to the tunnel coupling (non-interacting value $t$). The propagator $G^{\text{eq}}(i\omega)$ on the right-hand sides of Eqs.~(\ref{EQ:PT_Matsubara_dd})--(\ref{EQ:PT_Matsubara_(1l)(1l)}) is given by
\begin{eqnarray}
    \big[G^{\text{eq}}(i\omega)\big]^{-1}=
    \begin{pmatrix}
    i\omega-\epsilon  & t \\
    t  & i\omega-\Sigma^{\text{eq}}(i\omega) \\
    \end{pmatrix},
      \label{eq:nonmatrix}
\end{eqnarray}
with $\Sigma^{\text{eq}}(i\omega)$ being the Matsubara frequency lead self-energy on the second site of the SSH chain; the first site is treated explicitly as part of the interacting system. Here $\Sigma^{\text{eq}}(i\omega)$ is obtained from Eq.~(\ref{eq:leadselfenergy}) by the Wick rotation $\omega \rightarrow i\omega$ as well as $\kappa \rightarrow -\kappa$ in Eq.~(\ref{eq:boundary_gf_scaling_limit}). The integrations in  Eqs.~(\ref{EQ:PT_Matsubara_dd})--(\ref{EQ:PT_Matsubara_(1l)(1l)}) can be performed numerically. 

To illustrate that, in contrast to the non-interacting case, $W$ appears explicitly even if it is taken to be much larger than any other scale, we present the analytical result for the change of the tunnel coupling $\Sigma^{\text{int}}_{\text{d},1}$ for the case $\epsilon=0$, $\Gamma \ll \Delta \ll W$, and for a lead in its normal phase:
\begin{align}
\frac{\Sigma^{\text{int}}_{\text{d},1}}{t} = - \frac{2U}{\pi W} \ln{\left(\frac{ \Delta}{W} \right)} .
\label{eq:hoprenex}    
\end{align}
The explicit $W$ dependence visible here explains why we have to give the value of $W$ in Figs.~\ref{Fig:rho_int} and \ref{Fig:current_frg}.

To obtain the dot spectral function in first-order perturbation theory for the self-energy, we replace the single-particle parameters $\epsilon$, $V=0$, and $t$ on the right hand side of Eq.~(\ref{eq:nonmatrix}) by the ones complemented by the interacting self-energy $\epsilon+\Sigma^{\text{int}}_{\text{d,d}}$, $\Sigma^{\text{int}}_{1,1}$, and $t+\Sigma^{\text{int}}_{\text{d},1}$, invert the resulting matrix, perform the Wick rotation $i\omega \to \omega +i0$, and use Eq.~(\ref{EQ:dot_spectral_function}) for the (d,d) matrix element. The occupation follows from the spectral function via Eq.~(\ref{EQ:occupation}).

\subsection{Non-equilibrium steady state}

We next consider a setup with two SSH leads in the bias-voltage-driven non-equilibrium steady state and use the Keldysh formalism to obtain the self-energy to first order in $U$. 

We combine the first sites of the left and right lead with the dot site into an interacting subsystem. In the corresponding subspace the diagonal matrix elements as well as those associated with tunnel couplings of neighboring sites are non-vanishing. They are given as ($l=\mbox{L/R}$)
\begin{align}
    &\Sigma^{\text{int}}_{\text{d,d}}=\frac{-iU}{4\pi}\int^{\infty}_0 \hspace{-0.8em}d\omega_1 \int^{\infty}_{-\infty} \hspace{-0.8em}d\omega_2  \hspace{-0.4em} \sum_{l=\text{L/R}}  \hspace{-0.4em} S^{\text{K}}_{(l,1),(l,1)}(\omega_1,\omega_2),\label{EQ:PT_keldysh_dd}\\
    &\Sigma^{\text{int}}_{(l,1),(l,1)}=\frac{-iU}{4\pi}\int^{\infty}_0  \hspace{-0.8em} d\omega_1 \int^{\infty}_{-\infty} \hspace{-0.8em} d\omega_2  S^{\text{K}}_{\text{d,d}}(\omega_1,\omega_2) ,\label{EQ:PT_keldysh_r}
    \\ 
    &\Sigma^{\text{int}}_{\text{d,(R,1)}}=\frac{iU}{4\pi}\int^{\infty}_0\hspace{-0.8em} d\omega_1 \int^{\infty}_{-\infty}\hspace{-0.8em} d\omega_2  S^{\text{K}}_{\text{d,(R,1)}}(\omega_1,\omega_2) ,\label{EQ:PT_keldysh_d(1l)}\\
    &\Sigma^{\text{int}}_{\text{(L,1),d}}=\frac{iU}{4\pi}\int^{\infty}_0 \hspace{-0.8em}d\omega_1 \int^{\infty}_{-\infty}\hspace{-0.8em} d\omega_2  S^{\text{K}}_{\text{(L,1),d}}(\omega_1,\omega_2), \label{EQ:PT_keldysh_l}
\end{align}
with
\begin{align}
S^{\text{K}}(\omega_1,\omega_2)=-\partial_{\omega_1} \left[ \tilde{G}^{\text{R}}(\omega_1,\omega_2)\Sigma^{\text{K}}(\omega_1,\omega_2)\tilde{G}^{\text{A}}(\omega_1,\omega_2)\right]
\end{align}
and the superscript K stands for the Keldysh component. The non-vanishing matrix elements of  $\left[\tilde{G}^{\text{R}}(\omega_1,\omega_2)\right]^{-1}$ are 
\begin{align}
    &\left[\tilde{G}^{\text{R}}(\omega_1,\omega_2)\right]^{-1}_{(l,1),(l,1)}=i \omega_1+\omega_2-\Sigma^{\text{R}}_l(\omega_2)-V_l,\\
    &\left[\tilde{G}^{\text{R}}(\omega_1,\omega_2)\right]^{-1}_{\text{d},\text{d}}=i \omega_1+\omega_2-\epsilon,\\
    &\left[\tilde{G}^{\text{R}}(\omega_1,\omega_2)\right]^{-1}_{{\rm d},(l,1)}=\left[\tilde{G}^{\text{R}}(\omega_1,\omega_2)\right]^{-1}_{(l,1),{\rm d}}=t_l,
\end{align}
and $\left[\tilde{G}^{\text{A}}(\omega_1,\omega_2)\right]=\left[\tilde{G}^{\text{R}}(\omega_1,\omega_2)\right]^{\dagger}$. Moreover, the Keldysh component of the lead self-energy $\Sigma^{\text{K}}(\omega_1,\omega_2)$ is a diagonal matrix, and the corresponding matrix elements read
\begin{align}
    &[\Sigma^{\text{K}}(\omega_1,\omega_2)]_{(l,1),{(l,1)}}=-2\pi i \bar{\rho}_l(\omega_2)(t_{1}^l)^2\big[1-2f_l(\omega_2) \big] \nonumber\\
    &\hspace{10em}-2i\omega_1 \sgn(\omega_2),\\
    &[\Sigma^{\text{K}}(\omega_1,\omega_2)]_{\text{d,d}}=-2i\omega_1 \sgn(\omega_2),
\end{align}
where $f_l(\omega)$ is a Fermi function for the lead $l$ and $\bar{\rho}_l(\omega)$ is the boundary density of states of an SSH chain that has the same parameters as chain $l$, but with $t_1^l$ and $t_2^l$ exchanged.
More details can be found in Ref.~\onlinecite{Karrasch10}.

Finally, one can obtain the inverse of the effective retarded/advanced Green's function $G^{\text{R/A,int}}$ with the single-particle parameters changed by the interaction $\epsilon^{\text{int}}_{\text{d}} = \epsilon+\Sigma^{\text{int}}_{\text{d,d}}$, $\epsilon^{\text{int}}_{l} = \Sigma^{\text{int}}_{(l,1),(l,1)}$ and $t^{\text{int}}_l = t_l+\Sigma^{\text{int}}_{(l,1),(l,1)}$, as
\begin{widetext}
\begin{eqnarray} \label{Eq:eff_Green_R/A}
    \big[G^{\text{R/A,int}}(\omega)\big]^{-1}=
    \begin{pmatrix}
    \omega\pm i\eta-\epsilon^{\text{int}}_{\rm L}-\Sigma^{\text{R/A}}_{\text{L}}(\omega)  & t^{\text{int}}_{\text{L}} & 0 \\
    (t^{\text{int}}_{\text{L}})^{*}   & \omega\pm i\eta-\epsilon^{\text{int}}_{\text{d}} & t^{\text{int}}_{\text{R}} \\
    0 & (t^{\text{int}}_{\text{R}})^{*}  & \omega\pm i\eta-\epsilon^{\text{int}}_{\rm R}-\Sigma^{\text{R/A}}_{\text{R}}(\omega)
    \end{pmatrix}.
      \label{eq:nonmatrix2}
\end{eqnarray}
\end{widetext}

After the matrix Eq.~(\ref{eq:nonmatrix2}) has been inverted, the current can be computed employing Eq.~(\ref{EQ:AppCurrent}) but replacing the bare transmission coefficient by the effective one, i.e., the bare broadening and retarded/advanced Green's function in the first line of Eq.~(\ref{EQ:transmission_coefficient}) by the effective one,  $\Gamma_l(\omega)\rightarrow \Gamma^{\text{int}}_l(\omega)= -2\pi |t^{\text{int}}_l|^2 \text{Im}\big[G^{\text{R,int}}_{(l,1),(l,1)}(\omega)\big]/\pi$ and $G^{\text{R/A}}(\omega)\rightarrow G^{\text{R/A,int}}_{\text{d,d}}(\omega)$.

\subsection{The effective two-states picture}
In the case of a topological lead, one can, as for $U=0$, employ the intuitive, effective two-states picture to describe the effective dot spectral function and the occupation in first-order perturbation theory. We consider a single lead in its topological phase in equilibrium. W.l.o.g.\ we set $V_l = 0$ and suppress the lead index $l$ in the following.

As a new element compared to the non-interacting case, we have to consider a non-vanishing onsite energy on the first site of the lead given by $\epsilon_1^{\rm int} = \Sigma^{\text{int}}_{1,1}$.
As a consequence, the edge-state energy and its weight on the first site $\braket{1|\psi_{\rm e}}$ is changed. The latter enters the two-state calculation via the tunneling between the dot and the edge state. 
To find the effective edge state and its energy, we have to solve the SSH Hamiltonian, including $\epsilon^{\text{int}}_1$ as a boundary impurity.

We treat the first site as an impurity site with single-particle state $\ket{1}$ and energy $\epsilon^{\text{int}}_1$. This site is coupled with the hopping matrix element $-t_1$ to the first site (with corresponding state $\ket{2}$) of the remaining SSH chain, which starts with the hopping $-t_2$ between its first and second sites. It follows that this chain is in its trivial phase. We take the thermodynamic limit and refer to the single-particle Hamiltonian of this SSH chain as $\tilde h$. The total single-particle Hamiltonian of the system under consideration is then
\begin{align}
\bar{h} = \tilde{h} + \epsilon^{\text{int}}_1 \ket{1}\bra{1} - t_1 \big(\ket{2}\bra{1} + \ket{1}\bra{2}\big) .
\end{align}

We search for an effective edge state $\ket{\psi_{\rm e}^{\rm eff}}$ localized near the boundary that fulfills the Schrödinger equation $\bar{h}\ket{\psi_{\rm e}^{\rm eff}} = \epsilon_{\rm e}^{\rm eff}\ket{\psi_{\rm e}^{\rm eff}}$. We make the following ansatz
\begin{align}
\ket{\psi_{\rm e}^{\rm eff}} &= A \ket{\tilde{\psi}_{\pi + i\xi}} + c_1 \ket{1},\\
\ket{\tilde{\psi}_{\pi + i\xi}} &= \sum_{n = 1} ^{\infty} \Big[- \epsilon_{\rm e}^{\rm eff}e^{i(\pi + i\xi)n}\ket{2n} \nonumber\\
&\hspace{30pt} + \left(t_1 - t_2 e^{\xi}\right)e^{i(\pi + i\xi)(n+1)}\ket{2n+1}  \Big], \label{EQ:bulkstate}
\end{align}
with $\xi>0$ for normalizability. 
The ansatz has been made with a piecewise solution of the Schrödinger equation in mind. The state $\ket{\tilde{\psi}_{\pi + i\xi}}$ is the (unnormalized) projection of the most general form of an SSH edge-state wavefunction onto the chain starting at site $\ket{2}$. This directly gives us the energy up to a sign; we just need to insert $k = \pi + i \xi$ into the SSH dispersion $\epsilon_k = \pm \sqrt{t_1^2 + t_2^2 +2t_1t_2 \cos(k)}$. Squaring on both sides, we obtain:
\begin{align}
\left(\epsilon_{\rm e}^{\rm eff}\right)^2 =  t_1^2  +  t_2^2 -t_1t_2\left(e^\xi + e^{-\xi}\right). \label{EQ:ESenergy}
\end{align}
In the following, we now have to solve for $\xi$, or equivalently $e^{-\xi}$. Inserting our Ansatz into the Schrödinger equation leads to the two equations
\begin{align}
A\left(t_1 e^{-\xi} - t_2\right) +c_1 &= 0 ,\label{EQ:impESeq1}\\
-At_1\epsilon_{\rm e}^{\rm eff}e^{-\xi} + c_1\epsilon^{\text{int}}_1 &= c_1 \epsilon_{\rm e}^{\rm eff}\label{EQ:impESeq2}.
\end{align}
After a lengthy but straightforward calculation one obtains the following solution \cite{FN5}:
\begin{align}
e^{-\xi} &=  \frac{1}{2}\left(\frac{t_2}{t_1} - \frac{t_2^3}{t_1(\epsilon^{\text{int}}_1)^2}\right) \nonumber\\
&+
\sqrt{\frac{1}{4}\left(\frac{t_2}{t_1} - \frac{t_2^3}{t_1(\epsilon^{\text{int}}_1)^2}\right)^2 + \frac{t_2^2}{(\epsilon^{\text{int}}_1)^2}} . \label{EQ:expkappa}
\end{align}
The condition $\xi > 0$ (or equivalently $e^{-\xi}<1$) requires $|\epsilon^{\text{int}}_1|<t_2$; otherwise, there is no bound-state solution.
The effective edge-state energy can be obtained most easily from Eqs. (\ref{EQ:impESeq1}) and (\ref{EQ:impESeq2}). It reads
\begin{align}
\epsilon_{\rm e}^{\rm eff}= \frac{\epsilon^{\text{int}}_1}{2} +\frac{t_2^2}{2\epsilon^{\text{int}}_1} -\sgn(\epsilon^{\text{int}}_1)\sqrt{\frac{1}{4}\left(\epsilon^{\text{int}}_1 - \frac{t_2^2}{\epsilon^{\text{int}}_1}\right)^2 + t_1^2}.
\end{align}

From the normalization of the effective edge state, $\braket{\psi_{\rm e}^{\rm eff}|\psi_{\rm e}^{\rm eff}} \stackrel{!}{=} 1$, we can find its weight on the first site
\begin{align}
|\braket{\psi_{\rm e}^{\rm eff}|1}|^2 = |c_1|^2 = \frac{(1 - e^{-2\xi})(t_1-t_2e^{\xi})^2}{(\epsilon^{\text{eff}}_{\rm e})^2 + (t_1-t_2e^{\xi})^2}  . \label{EQ:ESweight}
\end{align}

Finally, the effective two-states Hamiltonian is given as 
\begin{align}
h^{\text{eff}}_{\text{two-states}} = \begin{pmatrix}
\epsilon^{\text{int}}_{\text{d}} & -(t^{\text{int}})^*\braket{1|\psi^{\text{eff}}_{\rm e}} \\
-t^{\text{int}}\braket{\psi^{\text{eff}}_{\rm e}|1}  & \epsilon^{\text{eff}}_{\rm e}
\end{pmatrix} \label{EQ:2stateH_eff} .
\end{align}
One can easily diagonalize Eq.\ (\ref{EQ:2stateH_eff}) to obtain the two effective eigenstates $\ket{\lambda^{\text{eff}}_{\pm}}$ with eigenenergy $\lambda^{\text{eff}}_{\pm}$. The corresponding effective two-states spectral function and effective dot occupation are given as
\begin{align}
\rho^{\text{eff}}_{\text{two-states}}(\omega)=&|\braket{\psi_{\rm d}|\lambda^{\text{eff}}_+}|^2 \delta(\omega-\lambda^{\text{eff}}_+)\nonumber\\&+|\braket{\psi_{\rm d}|\lambda^{\text{eff}}_-}|^2 \delta(\omega-\lambda^{\text{eff}}_-)
 \label{EQ:rho_frg_eff_two_level} ,
\end{align}
and
\begin{align}
n=\int^{0}_{-\infty} d\omega \rho^{\text{eff}}_{\text{two-states}}(\omega)
 \label{EQ:occupation_frg_eff_two_level} ,
\end{align}
respectively.


\begin{thebibliography}{}
\bibitem{FN1} The error made due to this approximation is $\mathcal{O}(\Gamma^2/\Delta)$ for peaks in the band region and $\mathcal{O}(\Gamma \sqrt{\Gamma/\Delta})$ for peaks in the gap. As $\Gamma \ll \Delta$, we conclude that the approximation is justified.
\bibitem{FN2} To be precise, one has to check if  ${\rm Re}\, \Sigma^{\rm R}(\tilde{\epsilon})$ varies strongly anywhere to make sure that $\tilde{\epsilon}$ does not intersect $\epsilon + {\rm Re}\, \Sigma^{\rm R}(\tilde{\epsilon})$ twice. Indeed, when $|\tilde{\epsilon}|$ approaches $\Delta$ from below, the derivative diverges like $-(\Delta - |\tilde{\epsilon}|)^{-\nicefrac{1}{2}}$. But when $|\tilde{\epsilon}|> \Delta$, ${\rm Re}\, \Sigma^{\rm R}(\tilde{\epsilon})$ is given by $\kappa\Gamma\Delta/(2\tilde{\epsilon})$, which varies slowly, so that $\tilde{\epsilon}$ and $\epsilon + {\rm Re}\, \Sigma^{\rm R}(\tilde{\epsilon})$ cannot intersect twice even here. There is indeed only a single solution.
\bibitem{FN3} Regarding the subtlety close to the band edges ($|\tilde{\epsilon}|\rightarrow \Delta$) raised in footnote \cite{FN2}, the analysis made there applies to a topological lead as well, i.e., there can only be a single peak solution close to a band edge.
\bibitem{FN4} Iterating to higher orders, we see that the series is perturbative in $\Gamma/\Delta \ll 1$ at the relevant voltages $V/2 = \Delta + |\tilde{\epsilon}|$ where the peak in $\rho(\omega)$ enters the band overlap region, so that truncating in lowest order is justified.
\bibitem{FN5} Technically, there are two more solutions, but these do not correspond to real, positive $\xi$.
\end{thebibliography}
\end{document}